# AN ALTERNATIVE USE OF THE VERDOORN LAW AT THE PORTUGUESE NUTS II LEVEL

**Vitor João Pereira Domingues Martinho**

Unidade de I&D do Instituto Politécnico de Viseu
Av. Cor. José Maria Vale de Andrade
Campus Politécnico
3504 - 510 Viseu
**(PORTUGAL)**
e-mail: vdmartinho@esav.ipv.pt

**ABSTRACT**

With this study we want to test the validity of the well known "Verdoorn´s Law" which considers the relationship between the growth of productivity and output in the case of the Portuguese economy at a regional and sectoral levels (NUTs II) for the period 1995-1999. The importance of some additional variables in the original specification of Verdoorn´s Law is also tested, such as, trade flows, capital accumulation and labour concentration. The main objective of the study is to confirm the presence of economies to scale that characterise the polarisation process with cumulative causation properties, explaining regional divergence. By introducing new variables to the original specification of Verdoorn´s Law we intend to examine how the economies to scale are influenced by the consideration of factors related to the Polarisation (Keynensian tradition) and Agglomeration (spatial economics tradition) phenomena. The results obtained from the regression analysis based on panel estimation show that the original specification of Verdoorn´s Law is more robust and confirm the presence of increasing economies to scale at both, regional and sectoral levels. However, the additional variables related to trade flows, capital accumulation and labour concentration have few influence on the performance of economies to scale (1)(Martinho, 2011).

## 1. INTRODUCTION

Several authors have developed a body of work in order to analyze the phenomenon of polarization (2)(Martinho e Soukiazis, 2005). Authors who have discussed the study of this phenomenon are mainly those associated with Keynesian theory, where forces of demand differences in explaining differences in regional growth. In models Keynesian tradition (3)Myrdal (1957), (4)Hirschman (1958), (5-7)Kaldor (1966, 1970 and 1981), among others), the polarization is based on growth processes with circular and cumulative causes, where export growth is the engine of regional growth, creating conditions for greater exploitation of economies of scale. In this process Verdoorn's Law is fundamental, since it guarantees the existence of scale economies growing, essential for growth processes occur with circular and cumulative causes.

The present study aims to examine several alternative specifications of Verdoorn's Law for each of the economic sectors of the Portuguese regions (NUTS II) and for each of these regions in the period 1995-1999. To do so, will present the equations of Verdoorn, Kaldor and Rowthorn (estimating Verdoorn's equation only for the reasons given later in this work) on the one hand, in its original form and the other with new variables in each equation. These variables are the ratio of GFCF/output (such as "proxy" for the accumulation of capital, given the lack of data to the stock of capital, by regions and sectors in the period considered), the ratio of the flow of goods/output and a variable that measures the level of concentration of population and economic activity. Alternatively to the variable concentration, was used, initially, the location quotient, typically used in the literature of Regional Economics, which compares the weight of the sectoral employment of a region with the weight of employment in the same region to the total national employment. However, it was chosen in this work, by rescinding its use at the expense of variable concentration, given the fact that the results obtained in the various estimates are less satisfactory than expected given the theory. The capital despite not having been considered in the original equations relating to the Law of Verdoorn, was later introduced by (8)Thirlwall (1980) and tested, for example, by (9)Leon-Ledesma (1998) for the Spanish regions.

The flow of goods is a variable widely used in the clustering models associated with authors such as (10)Krugman (1991), (11)Fujita et al. (1999 and 2000) and (12)Venables (1999), as a proxy for transport costs. As such, it seemed important to test the importance of this variable in the models of bias, since both processes are based on phenomena with circular and cumulative causes and the presence of scale economies growing. Therefore, we considered the flow ratio of goods/output in an attempt to link the theories of polarization of the Keynesian tradition with the agglomeration of recent tradition associated with the Space Economy.

The third new variable you want to measure the level of concentration of population and economic activity, calculated as the ratio between the number of regional employees in a given sector and the number of national employees in this sector, is also a widely used variable in the models of agglomeration, namely by (13)Hanson (1998). It should be noted, however, that agglomeration economies implicit in this variable turn out to be one of the assumptions underlying the Verdoorn relationship (with circular and cumulative growth). Nevertheless, the formalization of the models associated with the Verdoorn relationship with macro-economic reasons, does not take into account the direct effect of this variable that appears in models of agglomeration with a microeconomic foundation.

## 2. ALTERNATIVE MODELS THAT CAPTURE ECONOMIES OF SCALE

Kaldor (1966) in their attempt to revitalize the Verdoorn Law presented the following relations and tested them in an analysis "cross section" between industrialized countries:

$$p_i = a + bq_i, \quad \text{Verdoorn law} \quad (1)$$

$$e_i = c + dq_i, \quad \text{Kaldor law} \quad (2)$$

where pi, qi and ei are the growth rates of labor productivity, output and employment, respectively, with pi = qi - ei. Since then, and d = c = (1-b), which shows that in practice the estimation of an equation can define the parameters of the other.

(14-15)Rowthorn (1975 and 1979) suggested an alternative specification. That is, if it is assumed that the rate of growth is constrained by the supply of labor (hypothesis of the neoclassical theory of externalities), then the proper way to test the Verdoorn Law is directly link productivity growth (or output) with employment, considering, well, employment growth is exogenous. Thus, the equations Rowthorn considered to test the scale economies are the following:

$$p_i = \lambda_1 + \varepsilon_1 e_i, \quad \text{first equation of Rowthorn} \quad (3)$$

$$q_i = \lambda_2 + \varepsilon_2 e_i, \quad \text{second equation of Rowthorn} \quad (4)$$

where $\lambda_2 = \lambda_1$ e $\varepsilon_2 = (1 + \varepsilon_1)$.

Taking the above into account, our interest is to test empirically the relationship Verdoorn for the Portuguese economy at regional and sectoral, in order to identify savings to scale. Therefore, below is also presented an alternative specification that will later be estimated and analyzed. This specification, as noted earlier, equation Verdoorn results presented before, but now adding the ratio of GFCF/output ratio the flow of goods/output and the variable concentration of labor. The purpose of this specification is to test in the various economic sectors of the Portuguese regions in the period 1995-1999, the importance of capital (built with technical progress), thus avoiding errors incomplete specification. Introducing the flow of goods and the variable concentration is intended to test the importance of spatial factors in determining the economies of scale. The fundamental goal turns out to be joining the forces of polarization and clustering of this specification. The increased Verdoorn relationship is as follows:

$$p_i = a_0 + a_1 q_i + a_2 (C_i / Q_i) + a_3 (F_i / Q_{ik}) + a_4 (E_i / E_n), \text{ increased Verdoorn equation} \quad (5)$$

This equation is estimated for each economic sectors and all sectors of the five NUTS II of Portugal, over five years (1995 to 1999) and after individually for each NUTS II, with data disaggregated by four economic sectors, over the same period of time.

In this equation the variables increased pi and qi represent the growth of productivity and output, respectively. The variable (Ci/Qi) represents the ratio of GFCF/output (such as "proxy" for the variation of the ratio capital/output that incorporates technological progress), (Fi/Qik) represents the ratio of the flow of goods/output and ( Ei/En) represents the variable concentration. C is GFCF, Q is the gross value added, F is the flow of goods out of each of the regions (reflecting regional exports) and E is employment. Indexes i and n represent each of the regions and the national total, respectively. The index k represents the total industry.

## 3. DATA ANALYSIS

Taking into account the variables related to the model of Verdoorn presented previously in its original form and increased, the availability of statistical information, we used the following data disaggregated at the regional and sectoral. Annual data for the period 1995 to 1999 corresponding to the five regions of mainland Portugal (NUTS II) for the different economic sectors and the total economy of these regions. These data were obtained from the INE (National Accounts 2003).

## 4. EMPIRICAL EVIDENCE OF VERDOORN'S LAW IN PORTUGAL

Analyzing the coefficients of each of the estimated equations with the two estimation methods considered (Table 1), to point out, now and in general, the values obtained with both methods have some similarities. For agriculture, it appears that the Verdoorn coefficient has an elasticity outside acceptable limits, since it is above unity.

At the industry level Verdoorn coefficient (with an elasticity between 0.957 and 0.964, respectively, for the method of fixed effects and random effects) indicates the existence of strong increasing returns to scale, as expected, in the face of that by Kaldor, the industry is the engine of growth showing strong gains in productivity. The manufacturing industry presents the values closer to those found by Kaldor to the Verdoorn coefficient (between 0.509 and 0.781, respectively, for the two methods) and statistical siginifical.

**Table 1:** Analysis of sectoral economies of scale in five NUTS II of Portugal Continental, for the period 1995-1999

| Agriculture | | | | | | | | | |
|---|---|---|---|---|---|---|---|---|---|
| | **M.E.** | **Const.** | **$q_i$** | **$C_i/Q_i$** | **$F_i/Q_{ik}$** | **$E_i/E_n$** | **DW** | **$R^2$** | **G.L.** |
| **Verdoorn** | DIF | | 1.112* (10.961) | 0.066 (0.177) | -0.153* (-2.283) | -0.717 (-0.295) | 1.901 | 0.945 | 11 |
| | GLS | 0.483* (2.597) | 1.117* (14.538) | -0.668 (-1.560) | -0.182* (-3.594) | 0.065 (0.152) | 2.501 | 0.945 | 9 |
| **Industry** | | | | | | | | | |
| | **M.E.** | **Const.** | **$q_i$** | **$C_i/Q_i$** | **$F_i/Q_{ik}$** | **$E_i/E_n$** | **DW** | **$R^2$** | **G.L.** |
| **Verdoorn** | DIF | | 0.957* (5.425) | 0.213* (2.303) | -0.001 (-0.041) | -4.787* (-2.506) | 2.195 | 0.930 | 11 |
| | GLS | -0.089 (-0.591) | 0.964* (3.620) | 0.217 (1.558) | -0.023 (-0.515) | 0.042 (0.135) | 2.818 | 0.909 | 9 |
| **Manufactured Industry** | | | | | | | | | |
| | **M.E.** | **Const.** | **$q_i$** | **$C_i/Q_i$** | **$F_i/Q_{ik}$** | **$E_i/E_n$** | **DW** | **$R^2$** | **G.L.** |
| **Verdoorn** | DIF | | 0.509* (3.403) | 0.230* (5.081) | -0.141* (-3.705) | -4.331 (-1.527) | 2.052 | 0.945 | 11 |
| | GLS | -0.074* (-3.264) | 0.781* (3.861) | 0.138* (2.276) | 0.025* (3.384) | 0.070* (2.206) | 2.325 | 0.968 | 9 |
| **Other Industries** | | | | | | | | | |
| | **M.E.** | **Const.** | **$q_i$** | **$C_i/Q_i$** | **$F_i/Q_{ik}$** | **$E_i/E_n$** | **DW** | **$R^2$** | **G.L.** |
| **Verdoorn** | DIF | | 0.826* (6.279) | 0.056 (0.459) | 0.056 (1.032) | -2.934 (-1.718) | 2.106 | 0.791 | 11 |
| | GLS | -0.072* (-2.323) | 0.897* (7.479) | 0.146 (1.091) | -1.218 (-0.669) | 0.033 (0.467) | 2.114 | 0.917 | 9 |
| **Services** | | | | | | | | | |
| | **M.E.** | **Const.** | **$q_i$** | **$C_i/Q_i$** | **$F_i/Q_{ik}$** | **$E_i/E_n$** | **DW** | **$R^2$** | **G.L.** |
| **Verdoorn** | DIF | | 1.021* (5.430) | -0.116* (-2.587) | -0.020 (-0.856) | -5.458** (-1.895) | 1.369 | 0.846 | 11 |
| | GLS | -1.590 (-0.734) | 1.084* (5.577) | -0.106* (-2.319) | -0.020 (-0.815) | -5.985** (-2.063) | 1.629 | 0.717 | 9 |
| **All Sectors** | | | | | | | | | |
| | **M.E.** | **Const.** | **$q_i$** | **$C_i/Q_i$** | **$F_i/Q_{ik}$** | **$E_i/E_n$** | **DW** | **$R^2$** | **G.L.** |
| **Verdoorn** | DIF | | 0.905* (4.298) | -0.342* (-4.872) | -0.090* (-4.430) | -3.102* (-2.178) | 1.402 | 0.919 | 11 |
| | GLS | 1.559 (1.675) | 0.859* (3.776) | -0.371* (-4.665) | -0.096* (-4.404) | -3.158* (-2.098) | 1.459 | 0.912 | 9 |

**Note: * Coefficient statistically significant at 5%, ** Coefficient statistically significant at the 10% ME, estimation method, Const., Constant; Coef., Coefficient, GL, degrees of freedom; DIF method of estimation with fixed effects and variables in differences; GLS method of estimation with random effects.**

The other industry shows also strong increasing returns to scale through the Verdoorn coefficient. With respect to the coefficients of the new variables added, no statistical significance, reflecting little importance of these variables for the relationship between productivity growth and output growth in this sector. In the services sector the Verdoorn coefficient, although statistical significance is greater than one.

For the total regions, the Verdoorn equation presents results that confirm the existence of strong growing economies to scale, with additional variables to show statistical significance.

In a general analysis of Table 1, we verified the presence of strong economies of scale in the industry, confirming Kaldor's theory that this is the only sector with substantial gains in production efficiency.

Table 2 presents the results obtained in the estimations of these equations, but now the regional level.

By Verdoorn coefficients obtained for the different regions (NUTS II) of Portugal, it appears that increasing economies of scale are present only for the two estimation methods in the Norte, Lisboa e Vale do Tejo and the Algarve (since, in these regions, the Verdoorn coefficient is statistically significant and different from zero). Lisboa e Vale do Tejo has, as for the regression coefficient values significantly close to those obtained by Kaldor, showing, therefore, be a very important region in terms of scale economies growing in Portugal. The Centro and Alentejo regions also show some evidence, although less evident (as in one of the methods give values for the Verdoorn coefficient above unity), economies of scale increase.

**Table 2:** Analysis of scale economies at the regional level, 1995-1999

| Norte | | | | | | | | | |
|---|---|---|---|---|---|---|---|---|---|
| | **M.E.** | **Const.** | $q_i$ | $C_i/Q_i$ | $F_i/Q_{ik}$ | $E_i/E_n$ | **DW** | **$R^2$** | **G.L.** |
| **Verdoorn** | DIF | | 0.975* (9.996) | -0.051 (-0.223) | -0.097 (-0.684) | -3.585* (-2.194) | 1.945 | 0.968 | 8 |
| | GLS | 73.189 (0.225) | 0.982* (8.397) | -0.077 (-0.267) | -0.050 (-0.188) | -3.451 (-1.749) | 1.949 | 0.936 | 6 |
| **Centro** | | | | | | | | | |
| | **M.E.** | **Const.** | $q_i$ | $C_i/Q_i$ | $F_i/Q_{ik}$ | $E_i/E_n$ | **DW** | **$R^2$** | **G.L.** |
| **Verdoorn** | DIF | | 0.990* (8.945) | 0.102 (0.710) | 0.118 (0.818) | 0.396 (0.117) | 1.870 | 0.917 | 8 |
| | GLS | -0.503* (-2.728) | 1.067* (6.748) | 0.332* (3.943) | 0.227 (1.748) | 0.629* (3.292) | 1.937 | 0.957 | 11 |
| **Lisboa e Vale do Tejo** | | | | | | | | | |
| | **M.E.** | **Const.** | $q_i$ | $C_i/Q_i$ | $F_i/Q_{ik}$ | $E_i/E_n$ | **DW** | **$R^2$** | **G.L.** |
| **Verdoorn** | DIF | | 0.544* (2.561) | 1.017* (2.765) | -0.065 (-0.393) | 1.095 (0.413) | 1.734 | 0.858 | 8 |
| | GLS | 0.042 (0.226) | 0.674* (4.285) | 0.168 (1.490) | -0.006 (-0.034) | -0.203** (-1.927) | 1.930 | 0.725 | 11 |
| **Alentejo** | | | | | | | | | |
| | **M.E.** | **Const.** | $q_i$ | $C_i/Q_i$ | $F_i/Q_{ik}$ | $E_i/E_n$ | **DW** | **$R^2$** | **G.L.** |
| **Verdoorn** | DIF | | 1.026* (6.532) | 0.150* (2.816) | 0.092 (1.537) | -6.153 (-1.573) | 2.101 | 0.919 | 8 |
| | GLS | -0.109 (-1.544) | 0.971* (4.913) | 0.158* (2.857) | 0.015 (0.290) | 0.367** (2.090) | 1.734 | 0.986 | 11 |
| **Algarve** | | | | | | | | | |
| | **M.E.** | **Const.** | $q_i$ | $C_i/Q_i$ | $F_i/Q_{ik}$ | $E_i/E_n$ | **DW** | **$R^2$** | **G.L.** |
| **Verdoorn** | DIF | | 0.622* (3.303) | -0.080 (-0.597) | -0.133* (-4.752) | -17.050* (-4.270) | 2.230 | 0.890 | 8 |
| | GLS | 1.018* (6.667) | 0.626* (3.220) | -0.595* (-4.060) | -0.162* (-5.962) | -10.039* (-5.332) | 2.704 | 0.862 | 6 |

**Note: * Coefficient statistically significant at 5%, ** Coefficient statistically significant at the 10% ME, estimation method, Const., Constant; Coef., Coefficient, GL, degrees of freedom; DIF method of estimation with fixed effects and variables in differences; GLS method of estimation with random effects.**

## 5. FINAL CONCLUSIONS

In this study, we applied the Verdoorn Law to the sectoral Portuguese economy and at regional level, using panel estimations. The main objective is to identify the scale and significance to the economies of agglomeration forces in the regional and sectoral productivity.

In sectoral terms the equation Verdoorn increased shows during this period and for the NUTS II of Portugal, the existence of increasing returns to scale in industry, manufacturing, and in all other industrial sectors. In regional terms, this equation shows clear evidence of the existence of scale economies growing in the Norte, Lisboa e Vale do Tejo and the Algarve.

The consideration of new variables (GFCF ratio/output ratio flow of goods/output and the variable concentration) in equation Verdoorn little improvement in the Verdoorn coefficient in the estimations carried out with the original equation.

However, for these new variables should be noted that, in terms of sectoral concentration favors the variable productivity growth in manufacturing, which is the justification of the work associated with the New Economic Geography to give priority to this sector. On the other hand, capital relates negatively with productivity in services, indicating underutilized in terms of efficiency in this sector.

In regional terms, the Algarve capital relates also negatively correlated with productivity, possibly due to lack of investment quality. What was expected, given the importance of services in this region. The same factor has a positive influence on productivity growth in the Alentejo, Lisboa e Vale do Tejo and Centro. The flow of goods and variable concentrations have also negative values, with statistical significance, in the Algarve region. What would also be expected, given that the Algarve is a special area where the predominant sectors (services) do not produce tradable goods and not concentrate easily.

Finally, it should be noted that the Verdoorn coefficient captures much of the agglomeration effects and is therefore not necessary to express explicitly these effects.